\documentclass[journal=mamobx,manuscript=reprint]{achemso}

\usepackage{chemformula} 
\usepackage[T1]{fontenc} 
\usepackage{graphicx}

\usepackage{tikz}
\usepackage{color, amssymb}
\usetikzlibrary{shapes}

\newcommand{\markerlosangerouge}{\raisebox{0pt}{\tikz{\node[draw,scale=0.5,diamond,fill=red!10!red](){};}}}

\newcommand{\markcircblue}{\raisebox{0.5pt}{\tikz{\node[draw,scale=0.6,circle,fill=blue!20!blue](){};}}}

\newcommand{\markcirccyan}{\raisebox{0.5pt}{\tikz{\node[draw,scale=0.6,circle,fill=cyan!20!cyan](){};}}}

\newcommand{\markcircblack}{\raisebox{0.5pt}{\tikz{\node[draw,scale=0.6,circle,fill=black!20!black](){};}}}

\newcommand{\markertriorange}{\raisebox{0.5pt}{
		\tikz{\node[draw,scale=0.4,regular polygon, regular polygon sides=3,fill=orange!10!orange,rotate=0](){};}}}

\newcommand{\markertrimag}{\raisebox{0.5pt}{
		\tikz{\node[draw,scale=0.4,regular polygon, regular polygon sides=3,fill=magenta!10!magenta,rotate=0](){};}}}

\newcommand{\markertrigreen}{\raisebox{0.5pt}{
		\tikz{\node[draw,scale=0.4,regular polygon, regular polygon sides=3,fill=green!10!green,rotate=0](){};}}}

\newcommand{\markersquarewhite}{\raisebox{0.5pt}{
		\tikz{\node[draw,scale=0.6,regular polygon, regular polygon sides=4,fill=white!10!white,rotate=0](){};}}}	

\newcommand{\markersquaregray}{\raisebox{0.5pt}{
		\tikz{\node[draw,scale=0.6,regular polygon, regular polygon sides=4,fill=gray!10!gray,rotate=0](){};}}}

\author{Claire Schune}
\affiliation[SIMM]{Soft Matter Sciences and Engineering (SIMM), ESPCI Paris, PSL University, Sorbonne Universit\'e, CNRS, F-75005 Paris, France}

\alsoaffiliation[Solvay]{Solvay Silica, 15 rue Pierre Pays, BP 52, F-69660 Collonges-au-Mont-d'Or, France}
\author{Marc Yonger}
\alsoaffiliation[Solvay]{Solvay Silica, 15 rue Pierre Pays, BP 52, F-69660 Collonges-au-Mont-d'Or, France}
\author{Mohamed Hanafi}
\affiliation[SIMM]{Soft Matter Sciences and Engineering (SIMM), ESPCI Paris, PSL University, Sorbonne Universit\'e, CNRS, F-75005 Paris, France}

\author{J\"{u}rgen Thiel}
\affiliation[MPI]{Max-Planck Institute for Polymer Research, Ackermannweg 10, D-55128 Mainz, Germany}

\author{Laurent Guy}
\affiliation[Solvay]{Solvay Silica, 15 rue Pierre Pays, BP 52, F-69660 Collonges-au-Mont-d'Or, France}
\author{Thomas Chauss\'{e}e}
\affiliation[Solvay]{Solvay Silica, 15 rue Pierre Pays, BP 52, F-69660 Collonges-au-Mont-d'Or, France}
\author{Fran\c cois Lequeux}
\author{H\'{e}l\`{e}ne Montes}
\author{Emilie Verneuil}
\affiliation[SIMM]{Soft Matter Sciences and Engineering (SIMM), ESPCI Paris, PSL University, Sorbonne Universit\'e, CNRS, F-75005 Paris, France}
\email{emilie.verneuil@espci.fr}

\title[An \textsf{achemso} demo]
{Morphology and dynamics of dense nanometric precursor films of polymer melts}

\begin{document}
	
		


	
\begin{abstract}

Nanometer-thick supported films of polymer melts spontaneously form and spread around sessile droplets that are deposited on oxidized silicon wafers. At steady state, the films become dense and adopt a uniform thickness which is equal to twice the gyration radius of the free polymer. Remarkably, this law applies to a wide variety of melts and does not depend on the polymer chemistry nor on the surface state (oxide layer thickness, temperature, presence of water adsorbed, etc.). We show that existing theoretical descriptions cannot reproduce this experimental result. Conversely, the evolution toward this equilibrium state witnesses the specificity of the interactions at stake in these confined polymer films. The chains spreading dynamics can be modeled by taking into account both the polymer/surface friction and the polymer/polymer friction. 



  \end{abstract}


\section{Introduction}


Finely probing the interactions between melt polymer chains and a surface is of great relevance to any macroscopic mechanism involving adhesion, \cite{friddle_interpreting_2012} friction,\cite{singh_steady_2011, henot_friction_2018} or wetting \cite{PGG1985, venkatesh_interfacial_2022} of polymers on a solid. It is however quite challenging because of the intrinsic complexity of the physico-chemical mechanisms at stake, the possible coupling between surface and bulk effects, and the precision required for the measurements. Hence, experimental efforts have been made to probe the interfacial behavior\cite{zuo_conformational_2018} and among the methods developed in the past, a promising one is the approach based on wetting experiments of polymer melts on high energy surfaces, characterized by a positive spreading parameter $\mathcal{S}=\gamma_{sv}-(\gamma + \gamma_{sl})$, $\gamma_{sv}$, $\gamma$, and $\gamma_{sl}$ being the interfacial tensions of the solid/air, polymer/air, and solid/polymer interfaces, respectively. Indeed, when a polymer melt droplet is deposited on such surfaces, a so-called precursor film spreads around the sessile droplet as recently reviewed by Popescu et al. \cite{Popescu2012}. Precursor films have been evidenced a century ago \cite{Hardy1919}, but they have been precisely measured only over the past thirty years \cite{Daillant1988,Daillant1990,Daillant2000,Toney2000,Ausserre1986,Leger1988,Heslot1989bis,Heslot1989,Voue1998,Beaglehole1989,Cazabat1991,ViletteValignat1996,Valignat1993,Novotny1990,OConnor1996,OConnor1995,Mate2012,Albrecht1993,Albrecht1992,Schune2019,Schune2020,Silberzan1991,Esibov1998,Du2013,Moon2004,MullerBuschbaum1997,Seemann2001}. Their typical thickness being nanometric, they are extremely interesting systems to probe polymer/substrate interactions. The spreading of such thin films is thought of as an efficient way for the system to lower down its surface energy at early times.
In the literature \cite{Daillant1990, Silberzan1992}, it is observed that precursor films are comprised of two parts: (i) a submolecular part, which connects the uttermost end of the film to the bare substrate, and (ii) a thicker part connected to the droplet, which was sometimes shown to be dense \cite{Daillant1990}.
In a previous work \cite{Schune2020}, we studied the submolecular part of precursor films of polybutadiene melts. We showed that polymer molecules are in a non-dense state, and that the films spreading dynamics can be described by the sole friction of the polymer segments on the surface, which we measured. 
In the present study, we focus on the thicker part of precursor films, where polymer/surface interactions are expected to combine with polymer/polymer interactions.
The general theoretical framework to describe the spreading dynamics of dense polymer films relates the local thickness $h$ to its space and time derivative through Equation \ref{eq:eq_diff} \cite{Brochard1984,Churaev1982,Benzaquen_Salez_2014,Cormier2012}, where $x$ is the direction along the substrate - assuming no flux in the cross direction.
\begin{equation}
\frac{\partial h}{\partial t} =  - \frac{\partial}{\partial x} \left( \frac{\gamma}{3 \eta} h^{3} \frac{\partial^{3}h}{\partial x^{3}}\right) + \frac{\partial}{\partial x} \left( D(h) \frac{\partial h}{\partial x} \right) 
\label{eq:eq_diff}
\end{equation}
The first term on the right hand side arises from the capillary contribution to the pressure gradient, and the second one is a diffusive term. The diffusion coefficient $D(h)$ of the polymer chains can be written  \cite{mitlin_dewetting_1993, alizadeh_pahlavan_thin_2018} as the product of the mobility of the chains $M(h)$ with the derivative of the disjoining pressure $\partial \Pi(h)/ \partial h$. Depending on the system considered, $\Pi(h)$ may originate from a combination of van der Waals interactions, entropy, electrostatics, etc...
A few experimental studies reported on dense secondary films \cite{Daillant1990,Silberzan1992,ViletteValignat1996,Min1995,MullerBuschbaum1997,Esibov1998,Seemann2001, Mate2012, Jiang2016}, but no systematic experimental characterization have been performed to understand the origin of this film, mainly because the polymer systems used exhibited precursor films profiles with thickness variations extending over very narrow lateral length scales $dx$, making it challenging to analyze. In addition, no quantitative information on the polymer/surface interactions were extracted in a systematic way from the spreading dynamics of these films. 

Here, we offer to work with a series of  melts in the regime of pseudopartial wetting. This regime was first theoretically described \cite{Brochard:1991} and experimentally observed \cite{Silberzan1991} in the 1990's as the case where a sessile droplet at rest sits on its precursor film, and exhibits a non zero macroscopic contact angle while the precursor film possibly keeps spreading on the substrate. The aim of the present paper is two-fold: We will first describe the morphology of dense precursor films of polymers and discuss its thermodynamic or dynamic origin as suggested in the past \cite{Silberzan1991}. We will show that at steady state, dense precursor films exhibit a step-like profile of thickness proportional to the square root of the chains length, which is surprisingly independent of the polymer chemistry and the surface state. Second, we will characterize the chains spreading dynamics in the nanometer-thick films, and interpret it in terms of friction of the polymer segments with each other and with the surface.

\section{Materials and Methods}

The polymer melts used in this study are 1,4-polybutadiene (PBd 1,4), hydroxyl-terminated 1,4-polybutadiene (PBd-OH), 1,2-polybutadiene (PBd 1,2), hydroxyl-terminated polyisoprene (PI-OH), polystyrene (PS), and hydroxyl-terminated polystyrene (PS-OH). Molar masses $M_{n}$ range
between 400 and 40 000 g/mol and are given in Supplemental Material SM1 \cite{SM}, together with the Kuhn length $b$ of the polymers, the molar mass of the Kuhn segments $M_k$, and the polymolecularity indices IP. The experiments were performed in a hermetic cell under a weak flow of nitrogen, at a fixed relative humidity $RH_{c}$, which can be tuned between 11 and 91$\%$, and at different temperatures $T$ ranging between 10 to 73$^{\circ}$C. The precursor films spreading around the droplets were characterized with an ellipsometric microscope (EP3, Accurion) using nulling ellipsometry with 590 nm wavelength of 10 nm bandwidth, 65$^{\circ}$ incidence angle, and a 5 min time lapse between images. The pixel area over which measurements are averaged is 2 $\mathrm{\mu m^{2}}$.
Silicon wafers were either obtained from Siltronix (native oxide layer in the nanometer range) or from CEA-LETI (5 to 250~nm thick oxide layer obtained by dry oxidation). Their surface was carefully prepared: first, by immersion in a piranha solution (H$_2$SO$_4$ : H$_2$O$_2$ 2:1 v:v ) at 150$^o$C for 30 minutes, then rinsed in deionized water and dried with nitrogen in a clean environment. Prior to each experiment, a UV ozone treatment is applied for 30 minutes - unless otherwise stated. 
Sessile droplets of radius $r_{d}$ comprised between 40 and 250 $\mathrm{\mu}m$, and volume below the nanoliter, were deposited on these oxidized silicon wafers, with an initial oxide
layer thickness $e$ carefully measured and ranging from 2 to 250 nm depending on the wafer used. Noise reduction is achieved by taking advantage of the quasi circular shape of
the droplets: thicknesses are angularly averaged over 10$^{\circ}$ angles centered on the droplet center. Deposit method, experimental setup, and data analysis are available in previous
studies \cite{Schune2019,Schune2020}. 

\section{Results}
\begin{figure}[htpb]
	\centering
	\includegraphics[width= 17.2cm]{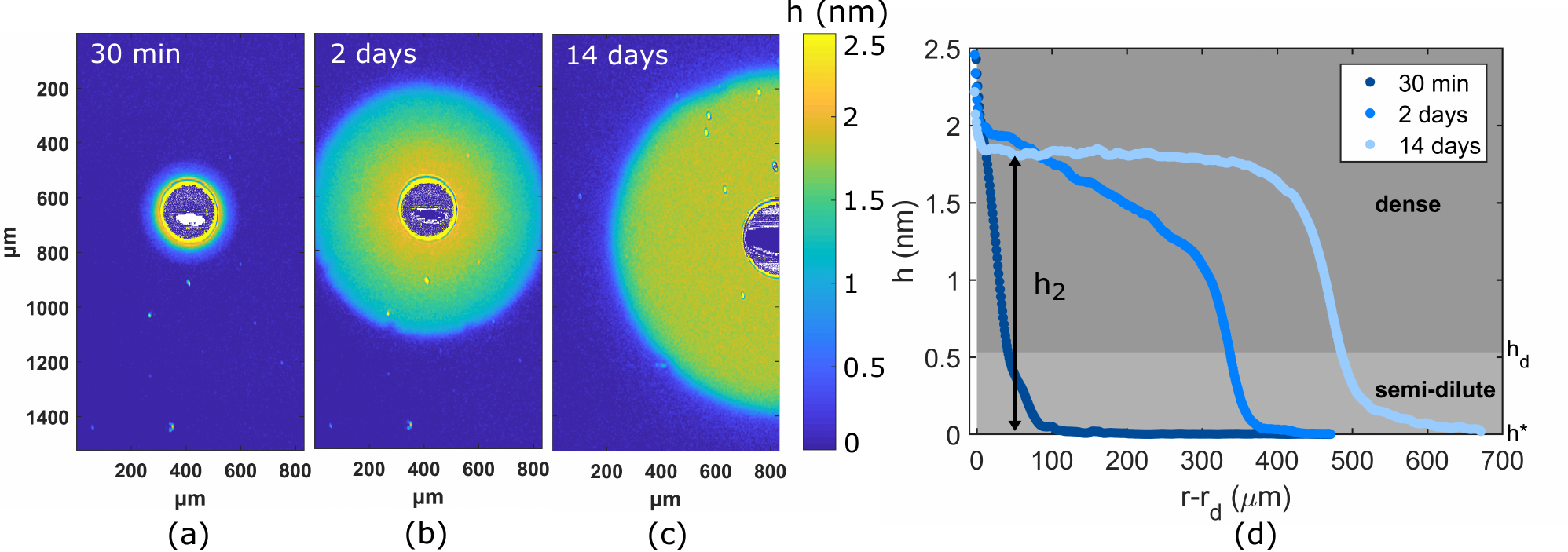}
	\caption{Thickness maps measured by ellipsometric microscopy around a sessile droplet (dark blue disk at the center) evidencing the spreading of a wetting precursor film. PI-OH with $M_{n}$ = 1200 g/mol. Images are taken at (a) 30 min, (b) 2 days, and (c) 14 days after deposit of the droplet. (d) Corresponding thickness profiles as a function of the distance to the droplet edge $r-r_{d}$, angularly averaged over a 10$^{\circ}$ angle centered at the droplet center. Thicknesses diverge at the edge of the droplet. Silica layer thickness: $e = 1.6$~nm; $T = 20^{\circ}$C; $RH_{c} = 11 \%$. }
	\label{fig:fig1}
\end{figure}
All the polymers used in the present study exhibit pseudo-partial wetting condition on oxidized silicon wafers: at rest, a sessile droplet coexists with a precursor film \cite{Brochard:1991, Silberzan1991,Schune2020,Popescu2012}.
The typical spreading of a precursor film is shown on Figure \ref{fig:fig1}. The dense film we observe here correspond to cases where the droplet as already stopped spreading on its precursor film, as shown in Fig.\ref{fig:fig1}a,b,c, while the film itself keeps on spreading.

\begin{figure}[htpb]
	\centering
	\includegraphics[width= 8.6cm]{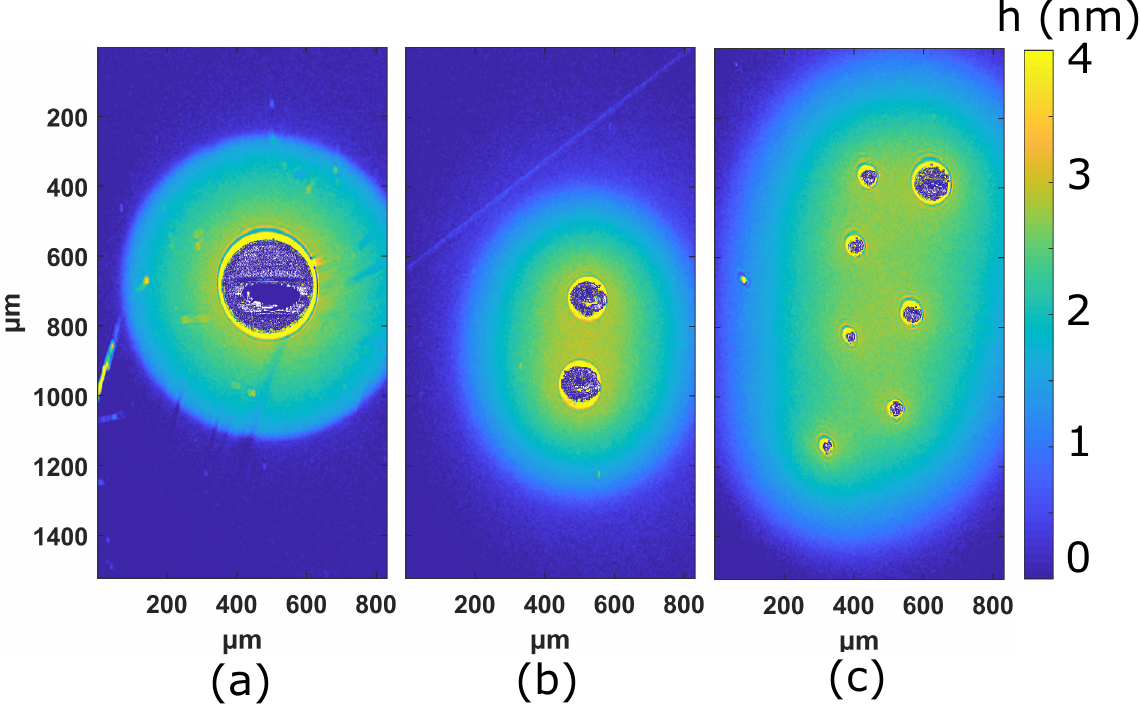}
	\caption{Thickness maps measured by ellipsometry of precursor films of PBd-OH of $M_{n}$ = 900~g/mol 8 hours after the droplets deposit, in the case of (a) one droplet, (b) two droplets, and (c) seven droplets. Silica layer thickness $e = 2.5$~nm, 1.6 nm and 1.3 nm, respectively. $T = 20^{\circ}$C; $RH_{c} = 11 \%$.}
	\label{fig:fig2}
\end{figure}

As detailed in a previous work, when the film thickness $h$ is lower than a molecular diameter $h_{d}$, the film is 2D semi-dilute: some silica sites are unoccupied  \cite{Schune2020,Semenov2003}. Below a thickness denoted $h^{\star}$, it is completely dilute: the chains do not interact with each other anymore. On the contrary, when $h>h_{d}$, it is dense. The dense part of the precursor films is called secondary film in the following.
As seen in Figure \ref{fig:fig1}(c,d), at long times the precursor film exhibits a step-like profile of uniform thickness $h_{2}>h_{d}$, which connects to the droplet. At the edge of the droplet,  measured thicknesses diverge, which allows to locate the droplet edge $r_{d}$. Furthermore, at earlier times (Fig.\ref{fig:fig1}(a,b), the connection between the film and the droplet also occurs at $h=h_2$, the plateau thickness (Fig. \ref{fig:fig1}(d)). Finally, if several droplets are deposited on the surface as in Fig.~\ref{fig:fig2}, precursor films eventually connect and the thickness in the inter-droplet area reaches a uniform value, which is also equal to $h_{2}$, the film thickness for a single droplet. Similar observations were made with all the polymers tested. These results suggest that $h_{2}$ is the steady state thickness of the film. It can equivalently be measured as the connection thickness of the secondary film to the droplet, or as the uniform thickness at long times. 

\begin{figure}[htpb]
	\includegraphics[width= 17cm]{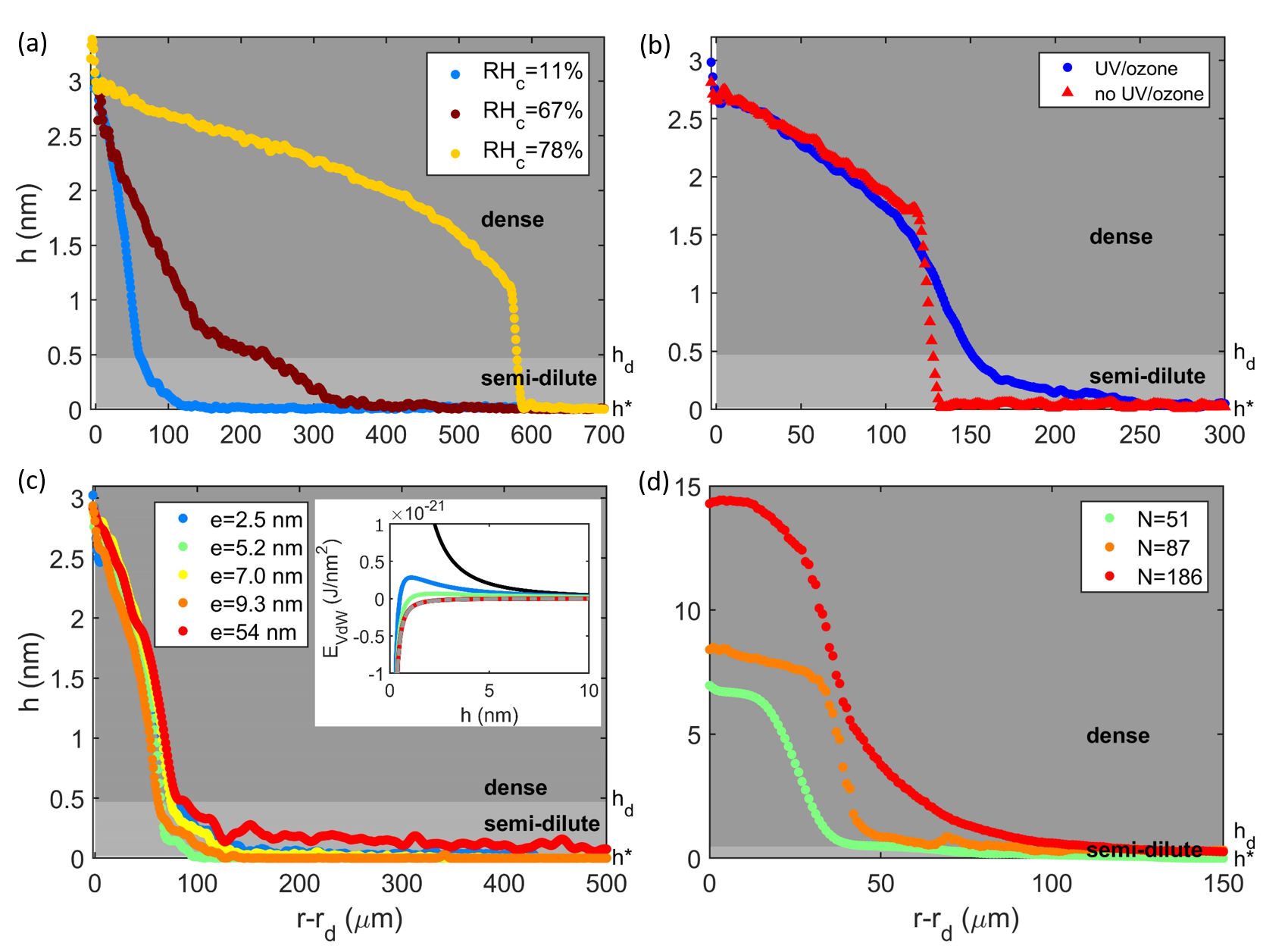}
	\caption{Thickness profiles of precursor films as a function of the distance to the droplet edge $r-r_{d}$. Data are angularly averaged over a 10$^{\circ}$ angle. (a) PBd-OH $M_{n}$ = 1300 g/mol, 30 min after the droplet deposit, at various humidity $RH_{c}$. T = 20$^{\circ}$C, 25$^{\circ}$C and 22.5$^{\circ}$C, respectively. (b) PBd-OH $M_{n}$ = 900 g/mol, with and without UV/ozone treatment of the wafer, 120 min after the deposit of the droplet. T = 20$^{\circ}$C; $e=2.5$~nm. (c) PBd-OH $M_{n}$ = 900 g/mol for various oxide layers $e$, 30 min after the deposit of the droplet. Inset: Van der Waals energy versus polymer film thickness $h$ computed from Eqs.1,2 in SM-2\cite{SM} for the different silica thicknesses $e$ (black $e=0$, dotted gray $e=\infty$). $T = 20^{\circ}$C and $RH_{c} = 11 \%$. (d) PBd 1,4; 1 day after the droplet deposit, for various chains lengths $N=M_{n}/M_{K}$ with $M_{K}=105$ g/mol. For experiments with longer chains, temperature was raised so that the radial extent of the film is similar : $r-r_d \sim 30~\mu$m : $T = 20^{\circ}$C for $N=51$, T = 47$^{\circ}$C for $N=87$ and T = 73$^{\circ}$C for $N=186$. $e=2.9$ nm, 2.5 nm and 3.2 nm, respectively. $RH_{c} \leq 11 \%$.
	}
	\label{fig:fig3}
\end{figure}

Interestingly, we observe that $h_{2}$ does not depend on the humidity in the cell (Figure \ref{fig:fig3}a), nor on the UV/ozone treatment applied to the wafer (Figure \ref{fig:fig3}(b). This thickness is therefore not sensitive to the exact physico-chemistry of the substrate. 

In addition, Figure \ref{fig:fig3}(c) 
shows that $h_{2}$ is independent of van der Waals interactions, which can be tuned by changing the oxide layer thickness $e$ \cite{Seemann2001}. On the other hand, the film profiles are modified when changing the polymer chain length: Figure \ref{fig:fig3}(d) 
presents precursor films profiles of PI-OH at long times for different chains lengths, $N$ being the number of Kuhn segments per chain ($N=M_{n}/M_{K}$).

\begin{figure}[htpb]
	\centering
		\includegraphics[width= 12cm]{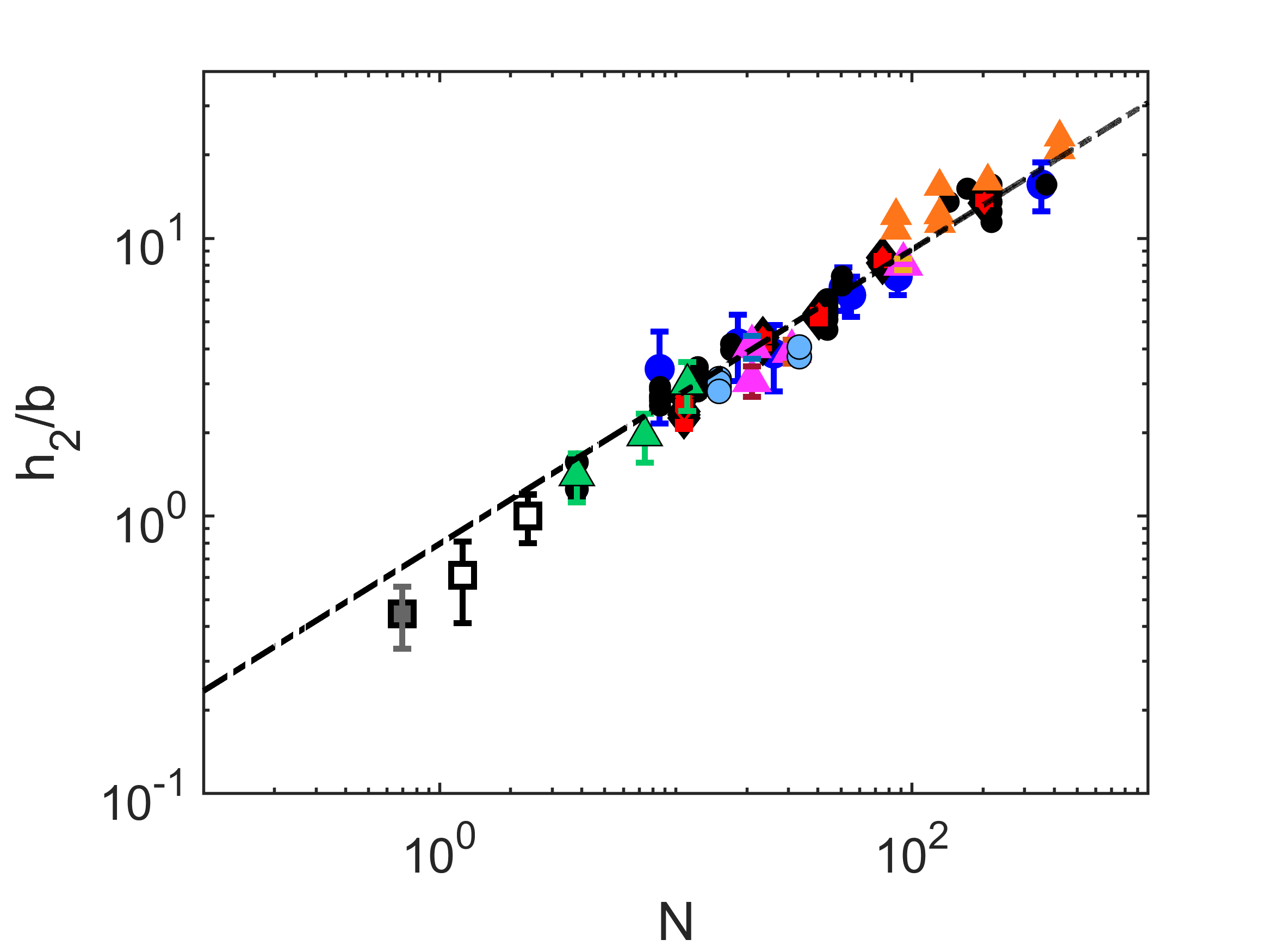}
		\includegraphics[width= 5.6cm]{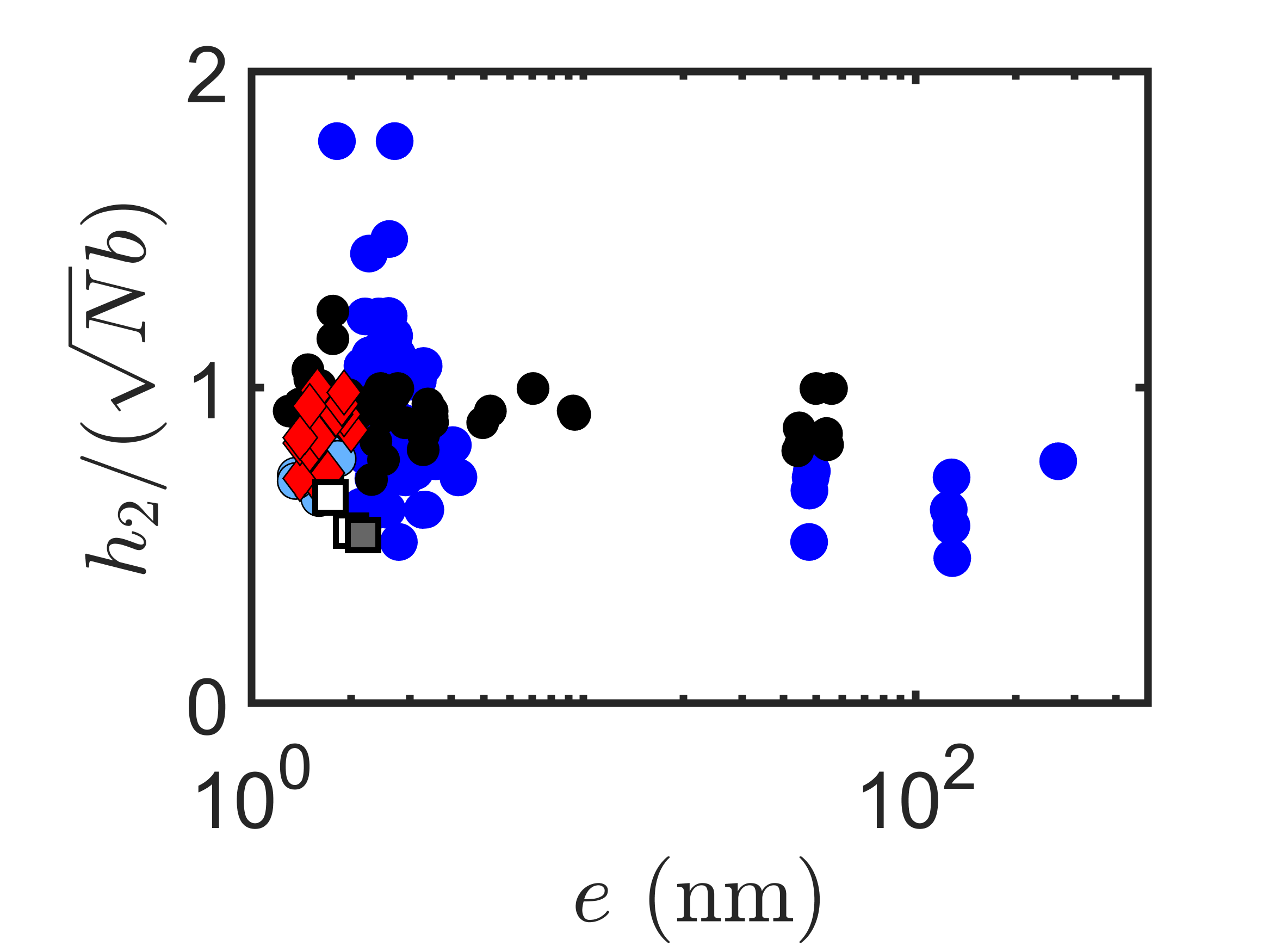}\includegraphics[width= 5.6cm]{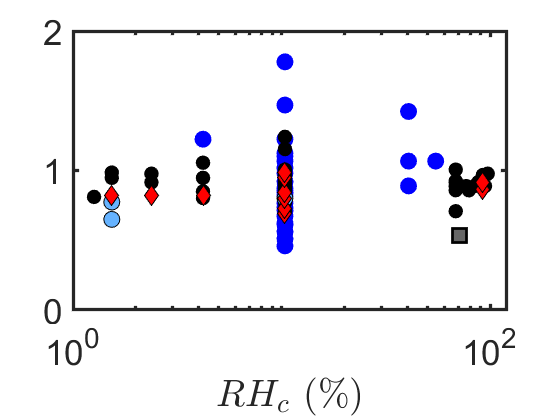}\includegraphics[width= 5.6cm]{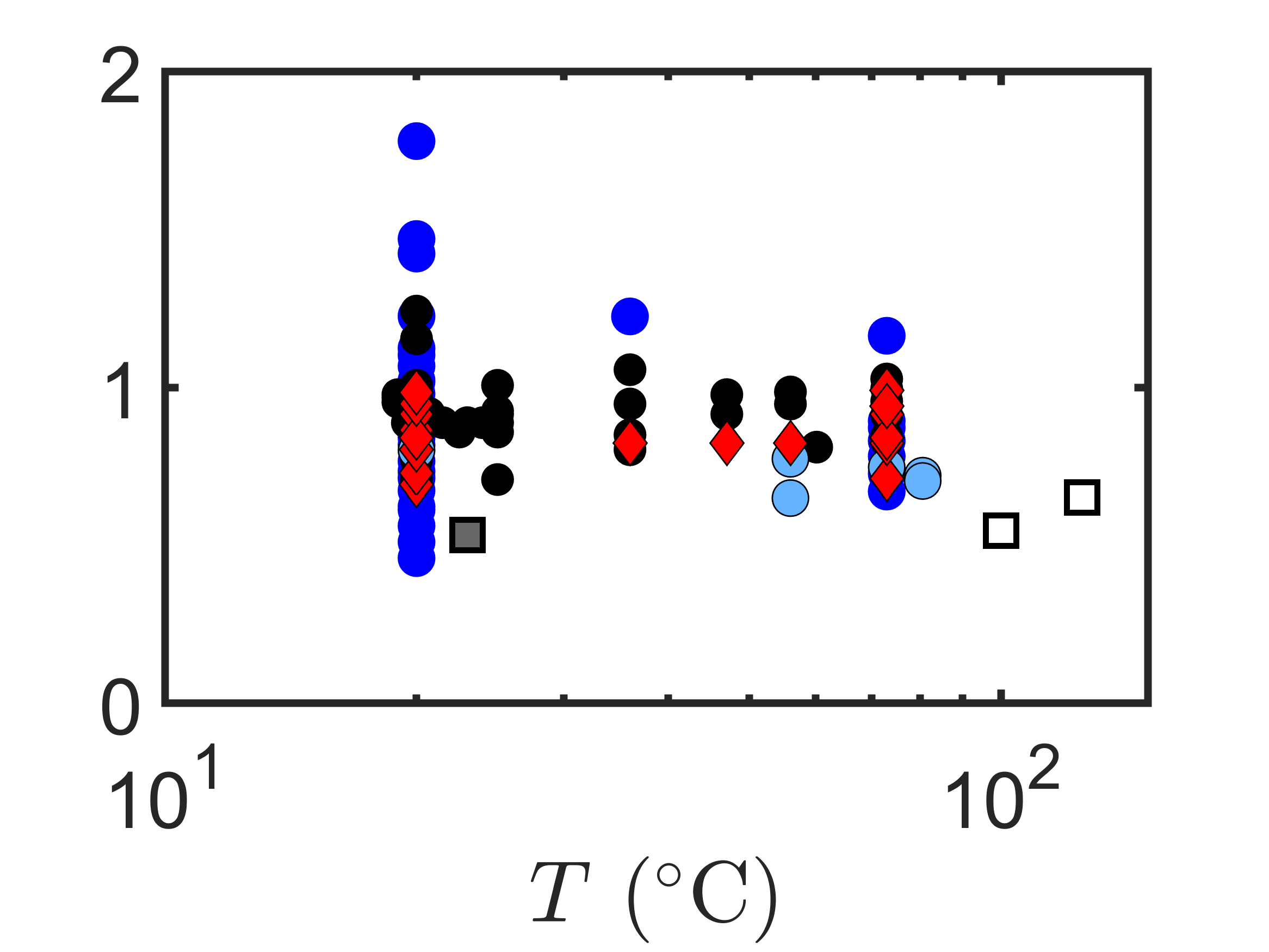}
	\caption{(Top) Secondary film thickness $h_2$ normalized by Kuhn length $b$ versus number of Kuhn segments $N$. Present study (averaged values): \protect \markcircblue  ~ PBd 1,4 \protect \markcirccyan ~ PBd 1,2 \protect \markcircblack ~ PBd-OH \protect \markerlosangerouge ~ PI-OH \protect \markersquaregray ~ PS \protect \markersquarewhite ~ PS-OH. Data from literature: \protect \markertriorange ~ PDMS from Silberzan {\it et al.} \cite{Silberzan1992}, \protect \markertrimag ~ PDMS-OH from Villette {\it et al.} \cite{ViletteValignat1996}             and Valignat {\it et al.} \cite{Valignat1993}, \protect \markertrigreen ~PFPE-OH from Min {\it et al.} \cite{Min1995}. Dashed line : $\mathrm{log}(h_{2}/ b) =p_{2}\mathrm{log}(N) + \mathrm{log}(p_{1})$ with $p_{2}=0.53$ et $\mathrm{log}(p_{1})=-0.23$. (Bottom) $h_{2}/(\sqrt{N}b)$ versus oxide layer thickness $e$, humidity $RH_{c}$, temperature $T$. }
	\label{fig:fig7}
\end{figure}

We measure that the longer the chains, the thicker the secondary films.
The thickness $h_{2}$ divided by the Kuhn length $b$ is plotted against the chains length $N$ on Figure \ref{fig:fig7}, for different polymers, various oxide layer thicknesses and relative humidities, and with or without a UV/ozone treatment of the wafer.
We observe that the whole data set collapses on a single master curve of equation $h_{2}/b = p_{1}N^{p_{2}}$, with $p_{2} = 0.53 \pm 0.02$ and $p_{1} = 0.79 \pm 0.07$, which holds for three decades in $N$.
In Figure \ref{fig:fig7}, we added the experimental data of Silberzan {\it et al.} \cite{SilberzanCRAS1991,Silberzan1992}, Villette {\it et al.} \cite{ViletteValignat1996}, Valignat {\it et al.} \cite{Valignat1993}, and Min {\it et al.} \cite{Min1995} who worked with PDMS, PDMS-OH and fluoropolymers, respectively, all these polymers exhibiting total wetting condition on oxidized silicon wafers instead of pseudo-partial wetting. This data nicely align with ours, which highlights the universality of the scaling law we measured. The power law dependence of $h_2$ with $N$ is strikingly similar to that expected for the radius of gyration:\cite{Colby2003} $R_g=1/\sqrt{6}b\sqrt{N}\sim 0.4b\sqrt{N}$. In the following, the thickness $h_2$ will be written as : 
\begin{equation}
    h_2\sim 0.8b\sqrt{N}
    \label{eq:h2:sqrt:N}
\end{equation}
Finally, the data from all our experiments can be compared by plotting $h_2/b\sqrt{N}$ as a function of the experimental conditions. This is done in Fig.~\ref{fig:fig7}, bottom part, where the silica thickness $e$, the ambient humidity and temperature were varied. These plots emphasize that none of these parameters change the film thickness, within experimental errors.\\

In the following, we focus on the spreading dynamics of the film, in light of Eq.~\ref{eq:eq_diff}. From our experimental thickness profiles in space and time, we first estimate the relative weight of the capillary contribution and the diffusive term in Eq.~\ref{eq:eq_diff}. The computation and the results are detailed in Supplemental Material SM-2 and SM-3 \cite{SM}. We find that the capillary term is always orders of magnitude smaller than the diffusive flux. 
In the following, we will consider that the capillary pressure do not contribute to the spreading of the film. Consequently, Equation \ref{eq:eq_diff} simplifies into: 
\begin{equation}
\frac{\partial h}{\partial t} = \frac{\partial}{\partial x} \left( D(h) \frac{\partial h}{\partial x} \right)
\label{eq:eq_diff_sanscap}
\end{equation}
The remaining diffusive term thus contains key information on secondary films. 

\begin{figure}[htpb]
	\centering
	\includegraphics[width= 17.2cm]{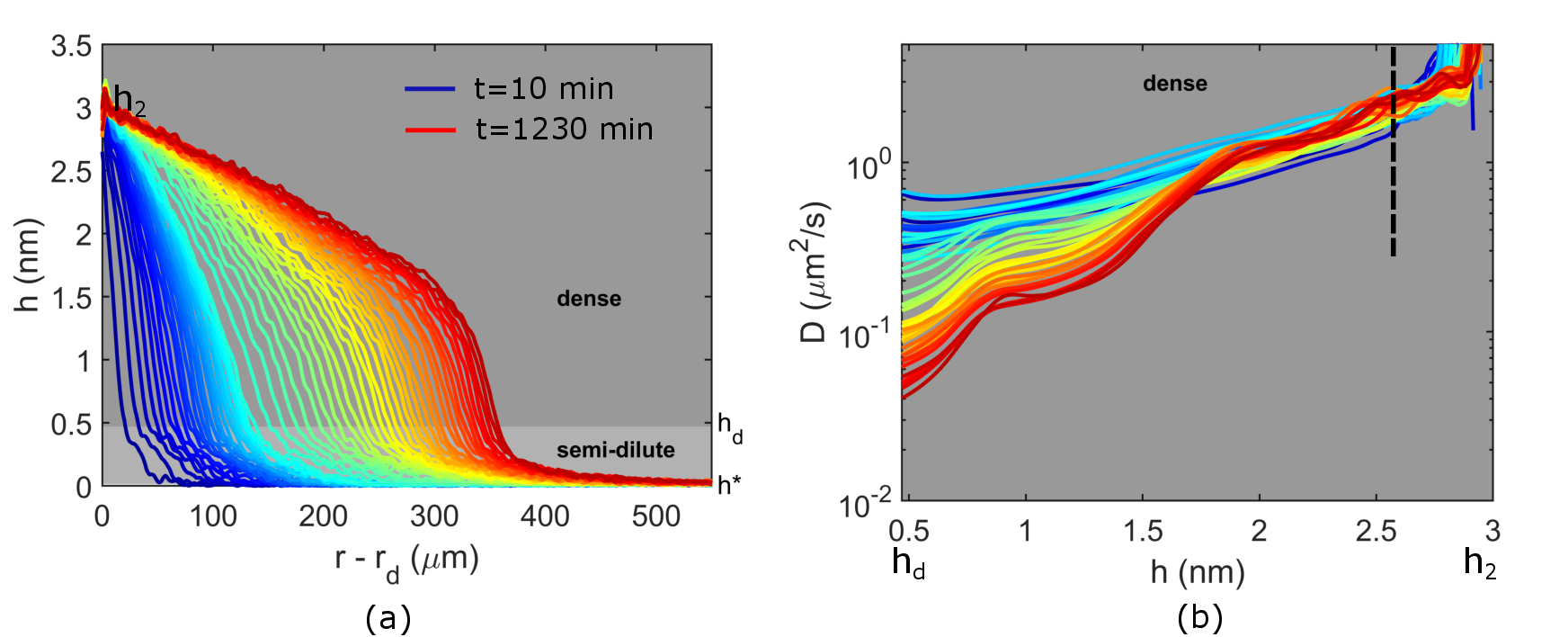}
	\caption{(a) Thickness profiles of a precursor film of PBd-OH 1300 g/mol at different times after the deposit of the droplet, comprised between 10 min (blue) and 1230 min (red). A time lapse of 5 min separates the curves until 120 min, then 30 min. $h_{2}=3$ nm. (b) Corresponding evolution of the diffusion coefficient $D$ with $h$ for the dense part of the film. $h_{2}$ and $h_{d}$ are indicated on the x-axis. Dotted line: threshold thickness $h_{th}=0.85 \times h_{2}$. $RH_c=11\%$. $T = 20^{\circ}$C.}
	\label{fig:fig8}
\end{figure}

By appropriate derivation and integration our experimental space-time thickness profiles, we can extract the variations of the diffusion coefficient with the film thickness $D(h)$ using Equation \ref{eq:eq_diff_sanscap}. Technical details on these computations are available in a previous paper \cite{Schune2020}.
Figure \ref{fig:fig8} presents thickness profiles with time of a secondary film of PBd-OH, and the corresponding $D(h)$ curves. 
We observe that $D$ increases with $h$. A temporal dependence of the diffusion coefficient is noted for $h \leq 0.7 \times h_{2}$. The range of thicknesses where $D$ presents a slowing down with time also corresponds to the part of the film that connects to the substrate, and which interestingly disappears when the surface is not activated with a UV/ozone treatment (see Figure \ref{fig:fig3}(b). Although of interest, this region is out of scope of this paper.
However, for higher thicknesses, $D(h)$ is time-independent.   
We chose to extract the diffusion coefficient at $h_{th}=0.85 \times h_{2}$, which we called $D_{S}$. This value of $h_{th}$ allows to be sufficiently close to $h_{2}$, while minimizing noise.

\begin{figure}[htpb]
	\centering
	\includegraphics[width= 17.2cm]{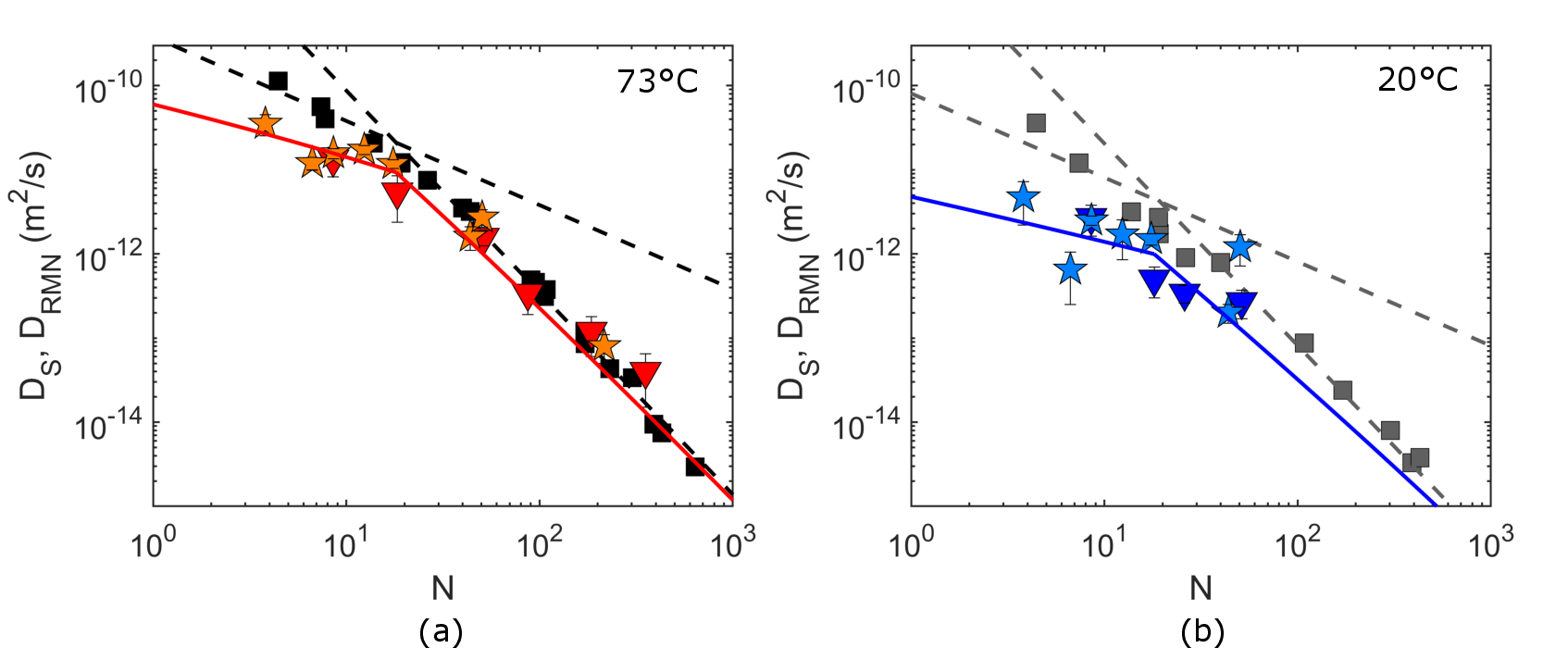}
	\caption{Diffusion coefficients versus chain length $N$. $RH_c$=11$\%$. (a) $T=73^{\circ}C$ (b) $20^{\circ}C$. ($\blacktriangledown$ $\bigstar$) Secondary precursor film data $D_{S}$.($\blacktriangledown$ ) PBd 1,4. ($\bigstar$) PBd-OH. ($\blacksquare$) Bulk NMR data from literature $D_{bulk}$ \cite{Guillermo2002,Lange2015,Meier2013}. Dashed lines: Bulk $D_{bulk}$ data are fit to Eqs \ref{eq:D_reptation_bulk} ($N>N_e$) and \ref{eq:D_bulk_Rouse} ($N<N_e$), with $N_{e}=18$, $\xi_{bulk}^{73^{\circ}C} = 1.4$ $10^{-11} ~ \mathrm{kg/s}$, $\xi_{bulk}^{20^{\circ}C} = 5$ $10^{-11} ~ \mathrm{kg/s}$. Lines: $D_s$ data are fit to Eq. \ref{eq:DS_Rouse_n} and Eq.\ref{eq:D_reptation_2}, using previously fitted value and $\xi_{surf}^{73^{\circ}C} = 1.4$ $10^{-10} ~ \mathrm{kg/s}$, $\xi_{surf}^{20^{\circ}C} = 1.3$ $10^{-9} ~ \mathrm{kg/s}$. }
		\label{fig:fig9}
\end{figure}

Figure \ref{fig:fig9} presents the evolution of $D_{S}$ with the chains length for PBd 1,4 and PBd-OH at $73^{\circ}$C (triangles and stars, respectively).
We measure that $D_{S}$ decreases with $N$. For $N<30$, the decrease is close to a power law $N^{-1}$. For higher values of $N$, the slowing down of the dynamics follows a higher exponent, as observed in Figure \ref{fig:fig9}(a). Furthermore, we observed that the presence of hydroxyl end-groups does not influence the dynamics.
This will be discussed in the following, and we will see that the spreading dynamics can be interpreted in terms of the sole friction of the polymer segments. 

\begin{figure}[htpb]
	\centering
	\includegraphics[width=8.5cm]{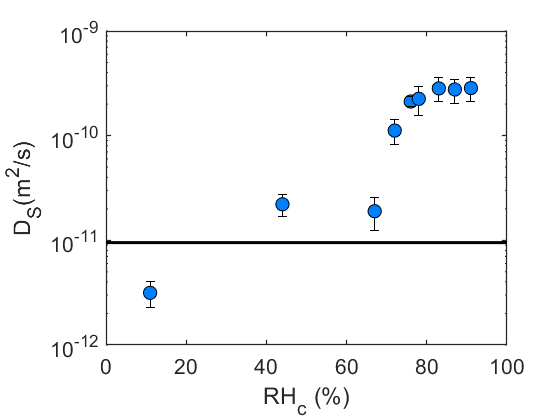}
 	\caption{
  Secondary film diffusion coefficient $D_{S}$ versus ambient relative humidity $RH_{c}$ \cite{Asay2005,Sumner2004,Takahagi2001,Sarfati2020}. PBd-OH 1300 g/mol ($N=12$). $T=22\pm 3^{\circ}C$. Black line: value of $D_{S}$ from Eq.~\ref{eq:DS_Rouse_n} with $\xi_{surf}=0$ and $\xi_{bulk}=5~10^{-11}$ kg/s. Above $RH_c\sim 70\%$, film spreading is faster than the zero friction case. 
  }
	\label{fig:fig10}
\end{figure}

While increasing the relative humidity in the cell does not influence the value of $h_{2}$ (Figure \ref{fig:fig3}(a)), it however greatly accelerates the spreading dynamics, as seen in Figure \ref{fig:fig10}.
We measured that the diffusion coefficient $D_{S}$ increases by two orders of magnitude when $RH_{c}$ varies from $11$ to $90\%$. This result holds for longer chains (measurements were performed for $N=8,12,50$) and whether chains are terminated by an hydroxyl group or not (data not shown). It was shown in the literature\cite{Asay2005,Sumner2004,Takahagi2001,Sarfati2020} that above a threshold relative humidity, water molecules adsorbed on oxidized silicon form a mobile layer thus modifying the surface state and the spreading velocity of polymers precursor films\cite{ViletteValignat1996}.

\section{Discussion \label{sec:discussion}}

Our measurements of the steady-state thickness of secondary films robustly highlight that $h_{2}$ only depends on the polymer chain length. This thickness is given by Eq.~\ref{eq:h2:sqrt:N}
or, equivalently, equals twice the gyration radius $R_{g}$ : $h_2\sim 2 R_g$. This law remarkably holds for polymers with varied chemistry. In addition, Equation~\ref{eq:h2:sqrt:N} is robust to changes in the experimental conditions such as temperature, physical chemistry of the substrate surface, as well as different regimes for spreading dynamics as will be detailed below.
To understand this universal scaling law, we offer to tackle the question of the origin of secondary films.
As mentioned previously, capillary effects are found negligible \cite{SM}. Energetic or dynamical effects are now considered.

\subsection{Is the film thickness set by the free energy of the system air/polymer film/silica/silicon?}

We first review the existing models based on thermodynamics that describe the variations with thickness $h$ of the free energy of thin and dense films of polymer melts supported by a solid substrate. Following pioneering descriptions\cite{Brochard:1991} of precursor films around droplets, our aim here is two-fold : first, determining whether the free energy of such systems exhibits a minimum at a given film thickness and, second, checking whether this thickness of minimum energy varies as the square root of the chain length. \\

For films of polymer melts supported by a solid in air, the free energy results from at least two contributions : (i) long range interactions and, in the present case, van der Waals interactions across the system interfaces (silicon/oxide, oxide/polymer, polymer/air) and (ii) the entropy of free polymer chains confined in a dense film. While the former does not depend on the polymer length $N$, the latter does, which makes it a good candidate to explain $h_2$ variations with $N$. However, in early theoretical works, De Gennes \cite{PGDG1987} derives the entropy of free polymers confined within a dense film by accounting for the screening of the repulsive interactions existing at small scale in melts. For such systems, De Gennes's model predicts that the entropy is independent of the film thickness for films thicker than a typical size. This size is the blob scale at rest, which can be taken as the Kuhn length $b$.In the past,the De Gennes's prediction was experimentally validated by surface force apparatus (SFA) measurements \cite{Kumar1998,Horn1988,Tirrell1989} : the force versus thickness curves were measured on melt polymer films between plates, and their modelling showed no contribution of the entropic term, demonstrating no thickness dependence of the entropy of confined free polymer chains.
Since almost all our films are thicker than $b$, no variations of entropy with thickness is expected to play a role in the selection of $h_2$ from De Gennes's prediction. 

 More recently, the De Gennes's vision was challenged by Lee and Johner \cite{Lee_Johner2019} who derived the free energy of confined polymer melts by random phase approximation (RPA). For thicknesses ranging between $b$ and $b\sqrt{N}$, the configuration entropy of non grafted polymer chains confined within a dense film is predicted to monotonously decrease with film thickness. The secondary films we measure mostly lay in this range since $h_2\sim 0.8 b \sqrt{N}$. In this model, the entropic contribution to the free energy is well approximated by $0.01 kT \times  h^{-2}e^{-h/(\sqrt{N}b)}$ for $h$ values in the vicinity of $h_2$ (note that this approximation does not hold when $h$ goes to zero). However, in the same range of thickness close to $h_2$, for both attractive and repulsive long range van der Waals interaction energy, we demonstrate in SM3 \cite{SM} that the combination of this entropic term to the van der Waals term yields a total free energy exhibiting no local minimum in the vicinity of $h_2$. Finally, for the sake of completeness, we note that non-monotonous variations of the entropic term have been predicted - and qualitatively evidenced - when a melt penetrates a {\it grafted} polymer layer, with a strong dependence on the number of loops and tails \cite{reiter_negative_2000,jopp_autophobic_1999,shull_wetting_1996,ferreira_scaling_1998,gay_wetting_1997,mensink_role_2021}. However, our experimental data show no evidence that polymer grafting occurs: no effect of the wafer cleaning or the end group on $h_2$ is observed. Hence, the polymers in the secondary films were assumed to be free.\\

 Altogether, the dense film thickness $h_2$ we measure, and particularly its robust variation as the square root of the polymer chain length, cannot be accounted for by available thermodynamics descriptions of non grafted polymer chains confined in dense supported films. Alternatively, we now examine whether the film thickness $h_2$ arises from dynamical effects, as offered by Bruinsma in the past \cite{Bruinsma1990}. In the following, we review the transport mechanisms at stake in the film spreading.

\subsection{Is the film thickness set by the dynamics of its spreading?}

The film we name here secondary films, of thickness $h_2$, spreads from the droplet edge, which is a reservoir of bulk liquid, toward a precursor film consisting of semi dilute non interacting polymer chains \cite{Schune2020}. It is dense. Here, we aim at characterizing both the driving and dissipative mechanisms controlling the spreading of such dense film. We first examine the possible dissipative mechanisms depending on the thickness.\\

In the past, different spreading regimes of liquid bodies on solid substrates have been thought of depending on the system size. We first consider macroscopic systems (droplets, thick films) with volumes far larger than a molecular volume. Their spreading can be successfully described within a hydrodynamic framework by considering a capillary driving term, viscous dissipation, and possibly dissipative mechanisms at the contact line which are not relevant here. Viscous dissipation arises because of the shear flow due to the assumed {vanishing} velocity at the substrate interface. However, this hydrodynamics description is expected to fail at the scale of the present secondary film. Indeed, thicknesses $h_2$ are measured equal to only two times the gyration radius of the chains: in such a confined regime, the details of the polymer chains movements at molecular scale cannot be reduced to a viscous term only. \\
In the literature, Johner and Obukov \cite{Johner2010} derive the flux of polymer chains within dense confined films of entangled polymers. In addition to the hydrodynamics flux described above, mobile chains diffuse by reptation within the network formed by the chains that have one or more monomers contacting the substrate, in a so-called permeation mechanism. They further demonstrate that the permeation flux is all the more efficient compared to the hydrodynamics flux that the film is thinner. More precisely, the viscous flux\cite{Mate2012} scales as $h^3/\eta$, where $\eta$ is the viscosity, while the permeation flux writes $kh$ where $k$ is a permeability which only depends on the polymer. From the derivation of the fluxes, a critical thickness $h_c$ is defined at which the spreading in the film switches from a viscous sheared regime to permeation, and writes $h_c=bN/N_e^{1/2}$ for entangled polymers. Note that in an earlier work, Bruinsma \cite{Bruinsma1990} also discusses this transition thickness, but arbitrarily sets $h_c$ at $R_g$.

The permeation description by Johner and Obukov provides a tool to determine the dissipative mechanism operating in the present secondary films, at least in the entangled regime (chains with length $N>N_e$). We first briefly discuss what entangled means in a quasi two dimensional film. Numerous theoretical works\cite{Brown1996,JLBarrat2018,Lee2017,Silberberg1981,Silberberg1988} predict a decrease of the entanglements density close to a surface, leading to an increase of the entanglement length $N_{e}^{2D}$ compared to the bulk value $N_e$. Applying these predictions \cite{Lee2017} to our systems, we find however that $N_{e}^{2D}$ remains close to the bulk value: it is not greater than $1.4 \times N_{e}$. For the sake of simplicity, we will consider here $N_{e}^{2D}=N_{e}$. In the following, we compare the critical thickness $h_c$ to the secondary film thickness we measure for polymer chains with length $N>N_e$. They both depend on $N$ so that we define a critical number of Kuhn segments $N_c$ for which $h_2<h_c$ if $N>N_c$. With $h_2\sim0.8\sqrt{N}b$, we find : $N_c\sim0.64 N_e$ so that $N_c$ is always smaller than $N_e$.   
In the entangled regime, we conclude that polymer chains in the secondary films are all in a permeation regime ($h_2<h_c$). It has been suggested in the past \cite{Bruinsma1990} that the secondary film thickness could be set by the transition between viscous and permeation regimes at the edge of the droplet. Hence, $h_2$ should vary in the same way as $h_c$, namely as $N$. This is not what we observe : $h_2$ scales as $\sqrt{N}$. This further demonstrates that $h_2$ is not set by the transition between viscous and permeation regime.

We will assume from now on that the flux in the dense thin films is limited by  permeation. Is $h_2$ controlled by the permeation flux as suggested in pioneering works?\cite{Bruinsma1990, Brochard1984}
Since precursor films exhibit a free interface with air, their thickness was assumed to adjust to the flux of polymers permeating through it. Our experimental data show that no change in the thickness $h_2$ was measured while very different fluxes could be obtained experimentally by either depositing several droplets or by tuning the relative humidity. More precisely, when several droplets are deposited as in Fig.~\ref{fig:fig2}, the flux in between them can be assumed to vanish, but the thickness remains the same as for spreading films.
 Moreover, experiments at high humidity lead to fast spreading (Fig.~\ref{fig:fig10}) but constant thickness, as shown in Fig.~\ref{fig:fig3}(a) and Fig.~\ref{fig:fig7}. Note that this last observation is also at variance with the description by Bruinsma which predicts a dependence of the film thickness on the slippage of the chains on the substrate. 
  
Finally, we examine the possible effect on $h_2$ of disjoining pressure gradients. Indeed, still in these pioneering theoretical works \cite{Bruinsma1990,Brochard1984}, assumption was made that the flux is driven by a gradient of disjoining pressure along the film length, and that the thickness depends on this gradient.
Again, our observations disagree with this prediction. Experiments where long range interactions were supposedly changed did not modify the film thickness. For example, changing the amplitude of the van der Waals interactions by changing the silica layer thickness $e$ (Fig.~\ref{fig:fig3}(c)~and Fig.\ref{fig:fig7}) did not change the flux nor the thickness. In addition, switching the sign of the van der Waals interactions does not affect the equation ruling the film thickness versus polymer size neither : while PB, PB-OH, PI, PI-OH and PS all bear attractive van der Waals interactions on oxidized silicon wafers (pseudo-partial wetting, conjoining Derjaguin pressure), PDMS or fluorinated polymers bear repulsive ones (total wetting, disjoining Derjaguin pressure). Therefore, we find no evidence of the effect of the disjoining pressure gradient on the film thickness value.\\

 In our understanding, these observations show that the available theoretical models fail to account for the film thickness of dense secondary precursor films measured by us and by others. We now turn to a discussion on the diffusion coefficient data.\\

\subsection{What can be learnt from the spreading dynamics of dense precursor films?}

\subsubsection{Entangled polymers $N>N_e$ at low ambient humidities}

We first focus on entangled polymer chains spreading at low ambient humidity. We measure a diffusion coefficient $D_S$ at a film thickness below but close to $h_2$ that characterizes the chain diffusion within the film. In Figure~\ref{fig:fig9}, $D_S$ is plotted as a function of the chain length $N$, together with bulk diffusion coefficients taken from the literature and measured in non confined melts at the same temperature (square symbols). Interestingly, $D_S$ and $D$ are found equal, within experimental uncertainty, for $N>N_e$.
We furthermore find that the diffusion coefficient in the films varies with chain length according to the power law with exponent -2.4 which characterizes the reptation mechanisms of entangled polymers: \cite{Colby2003}
\begin{equation}
D_{rep}^{bulk}=\frac{kT}{\xi_{bulk}}\frac{N_e^{1.4}}{N^{2.4} }
\label{eq:D_reptation_bulk}
\end{equation}
where $\xi_{bulk}$ is a friction coefficient between Kuhn segments. By analogy with polymer diffusion in the bulk, this suggests that, in films, the driving mechanism is also the gradient of position entropy of the chains along the film length. Altogether, the picture we offer is the following : for a film thickness below but close to $h_2$, the spreading of polymers in the film is driven by the position entropy gradient and limited by a permeation mechanism where friction mostly occurs between monomers with no significant account of the substrate/monomer friction. This striking result is somewhat unexpected given the system complexity.

In the following, we will show both monomer/monomer and substrate/monomer friction must instead be taken into account for films with lower thicknesses: this is the case for dense films of non entangled PBd and PB-OH polymers.

\subsubsection{Non entangled polymers $N<N_e$ at low ambient humidities}

For chain lengths below the entanglement length $N<N_e$, in bulk, polymer melts no longer diffuse by reptation. Instead, they obey a Rouse dynamics : the bulk diffusion coefficient is set by the friction of the whole chain, which is the sum of the friction experienced by each monomer or Kuhn segment. Assuming a homogeneous friction coefficient for all monomers equal to $\xi_{bulk}$, the polymer chain diffusion coefficient in bulk writes: \cite{Colby2003}
\begin{equation}
    D_{bulk}^{Rouse}=\frac{kT}{N \xi_{bulk}}
    \label{eq:D_bulk_Rouse}
\end{equation}
For dense secondary precursor films consisting of non-entangled polymers of PBd and PB-OH that bear a hydroxylated terminal group, we find a change in the exponent of the power law variation of $D_s$ with the chain length in the vicinity of $N_e$ (Fig.~\ref{fig:fig9}). Our data do not allow us to measure this exponent. However, we find that $D_s$ is significantly smaller than the bulk diffusion coefficient - at the same length and temperature, as opposed to the entangled regime discussed above where chain diffusion coefficients in bulk and in films were equal. This observation suggests either additional friction comes into play, or the driving mechanism for spreading is different for low mass polymers.  \\
From the observation that the film thickness $h_2$ varies as $0.8b\sqrt{N}$ whatever the chain length, we offer to assume that the driving mechanism for spreading in the film is the gradient of position entropy for all $N$s. In this framework, we suggest the friction mechanism at stake for low mass polymers consists of both monomer/monomer friction and monomer/substrate friction. This hypothesis was verified in a previous study: \cite{Schune2020} for non dense films with effective thickness below $b$, we showed monomer/substrate interactions were setting the diffusion coefficient of dilute or semi dilute PBd or PBd-OH polymers on silica. In dense films, low mass polymers exhibit low film thickness $h_2$ so that the ratio of the surface friction term to the chain/chain friction term becomes comparatively larger. In this framework, we suggest that the non entangled chains diffuse in the film according to a Rouse mechanism as in bulk, but with a friction coefficient that depends on the monomer location : for monomers interacting with the substrate, the friction coefficient is denoted $\xi_{surf}$ and for monomers interacting with other monomers, it is denoted $\xi_{bulk}$. We define $n$ as the number of monomers per chain interacting with the surface. As a result, the diffusion coefficient of chains in the dense film for polymers of length $N<N_e$ writes :
\begin{equation}
D_{S}^{Rouse} = \frac{kT}{n \xi_{surf} + (N-n)\xi_{bulk}}
\label{eq:DS_Rouse}
\end{equation}
The number $n$ can be estimated as follows. Let $h_d$ being a molecular diameter and $b$ the segment length, the volume $v$ and surface $\sigma$ of a monomer (or Kuhn segment) write: $v=bh_d^2$ and $\sigma=bh_d$, while $h_d$ can be written \cite{Schune2020} as a function of the polymer density $\rho$ and Avogadro number $N_A$ as  $h_d=\sqrt{\frac{M_k}{b\rho N_A}}$. In a dense film of thickness h and surface S, there is $\alpha$ polymer chains of length $N$ and $n_t$ monomers densely covering the surface. The volume
occupied by the chains writes $\alpha N v = hS$ with $S =n_t\sigma$. So $n =n_t/\alpha =Nh_d/h$.
Diffusion coefficients $D_s$ are measured at a thickness $h_{th}=0.85 h_2$. With $h_2=0.8b\sqrt{N}$, $n(h_{th})$ writes: $n=\sqrt{N}h_d/0.68b$ so that $D_S^{Rouse}$ becomes:
\begin{equation}
D_{S}^{Rouse} = \frac{kT}{N( \xi_{surf} \frac{h_{d}}{0.68b\sqrt{N}} + \xi_{bulk}(1 - \frac{h_{d}}{0.68b\sqrt{N}}))}
\label{eq:DS_Rouse_n}
\end{equation}
Equation~\ref{eq:DS_Rouse_n} was compared to the experimental data of $D_s$ versus $N$ in Figure~\ref{fig:fig9}, for $N<N_e$, with the surface friction coefficient $\xi_{surf}$ as a fitting parameter. We set the bulk friction coefficient $\xi_{bulk}$ between monomers at the value reported in the literature \cite{Guillermo2002,Meier2013,Lange2015}: $\xi_{bulk}=1.4\times 10^{-11}$~kg/s at $73^{\circ}$C and $\xi_{bulk}=5\times 10^{-11}$~kg/s at $20^{\circ}$C. For PBd and PBd-OH polymers, $b$=0.96~nm and $h_d=0.47$~nm.\cite{SM} Fitting Equation~\ref{eq:DS_Rouse_n} to the data (full lines in Fig.~\ref{fig:fig9} for $N<N_e$) gives :  $\xi_{surf}^{73^{\circ}C}=1.4\pm1 \times10^{-10}~ \mathrm{kg/s}$ and $\xi_{surf}^{20^{\circ}C}=13\pm5 \times10^{-10}~ \mathrm{kg/s}$.\\
We find that the monomer/substrate friction coefficient in about 10 times larger than the bulk friction coefficient at $73^{\circ}$C, and 30 times larger at $20^{\circ}$C. In addition, a numerical estimate of $n$, the number of monomer contacting the substrate, shows that for the PBd series with $N$ between 7 and 400, the chains always have at least 2 monomers at the surface. This is consistent with a permeation scenario.
From this result, we look back at the entangled case and account for the effect of surface friction by extending Eq.~\ref{eq:DS_Rouse_n} to the entangled situation simply using Equation~\ref{eq:D_reptation_2}: 
\begin{equation}
D_{S}^{rep}=D_S^{Rouse}\left(\frac{N_e}{N}\right)^{1.4}	
\label{eq:D_reptation_2}
\end{equation} 
The corresponding variation of the diffusion coefficient with the chain length $N$ for $N>N_e$ is plotted in plain lines in Fig.~\ref{fig:fig9}. 
The difference between the predictions given by Eq. \ref{eq:D_reptation_bulk} (dashed lines) and  Eq. \ref{eq:D_reptation_2} (full lines) lays within the range of experimental error bars. This confirms that monomer-monomer friction  and monomer-substrate friction cannot be discriminated on thick entangled films.

\subsubsection{High humidity}

Interestingly, we find that the diffusion coefficient of the chains in the dense films increases non linearly with ambient humidity (Fig.~\ref{fig:fig10}): above RH=60$\%$, the diffusion coefficient  
sharply increases by one order of magnitude, then plateaus for RH>80$\%$.
The enhanced mobility of adsorbed layers of water could be responsible for the strong acceleration of the spreading dynamics of polymer chains on highly hydrated silica. Similar results were obtained in the past for PDMS melts within porous packings of silica \cite{venkatesh_interfacial_2022} and for PDMS precursor films \cite{ViletteValignat1996}, for both methyl and hydroxyl-terminated PDMS. In the latter, a significant increase in precursor films diffusion coefficients was observed above 70\% RH and ascribed to a decrease in the friction coefficient between the molecular layer in contact with the substrate \cite{degennes_spreading_1990} and the more hydrated silica surface, rather than to a change in the interaction energy. Along this line, assumption is first made that these adsorbed water layers decrease the friction coefficient between monomer and substrate. In the limiting case were the surface friction vanishes, $\xi_{surf}=0$, chains would collectively slip on the surface into a "plug-flow" with no monomer-monomer friction neither. This is needed what we find: by setting $\xi_{surf}=0$ in Eq.~\ref{eq:DS_Rouse_n}, we consistently compute a diffusion coefficient value that is lower than the measured ones. It is plotted as a plain line in Fig.~\ref{fig:fig10} for $N=12$. Yet, the limiting and driving mechanisms controlling this fast spreading still lack a comprehensive description when adsorbed layers of water cover the silica surface.  Nevertheless, we emphasize again that the film thickness $h_2$ always varies with $N$ as $0.8 b \sqrt{N}$, even in this fast spreading regime. \\
 
\section{Conclusion}
In this paper, using a series of polymers with varied size and chemistry in a systematic way, we show that precursor films of polymer melts in pseudo-partial wetting condition comprise a dense part connecting the droplet to the semi-dilute part of the film. This dense part is called secondary film, and, at steady-state, exhibits a step-like profile of thickness proportional to the square root of the chains length. More precisely, this thickness is equal to twice the radius of gyration of the polymers. This universal law is measured over three decades in chains length and also applies to data from the literature obtained on polymers that were in total wetting conditions. Remarkably, it does not depend on the sign or the range of van der Waals interactions, nor on the chemical nature of the polymer chains or their end-functionalization, nor on the surface state of the substrate. {This robust experimental result challenges the existing theories, and invites to refine the current theoretical framework for polymers in precursor films.}\\
Nevertheless, from the measurements of the secondary film spreading dynamics, the following picture emerges: polymers are dynamically adsorbing to the surface by few monomers and form a transient network through which other polymers permeate \cite{Johner2010}, driven by a gradient of position entropy. Our data allow us to quantitatively discriminate between the friction of the polymer segments with each other and with the surface, the latter being 10 times larger on poorly hydrated surfaces. 
We hope our experimental findings will stimulate further theoretical developments.\\

https://doi.org/10.1021/acs.macromol.3c00554\\

This document is the unedited Author's version of a Submitted Work that was subsequently accepted for publication in Macromolecules, copyright \textcopyright American Chemical Society after peer review. To access the final edited and published work see:\\ https://pubs.acs.org/articlesonrequest/AOR-MTQ9KXC8EXPKR4IQ2XIX

\section*{Acknowledgments}
We deeply thank Albert Johner for numerous discussions. 

\section*{Supporting Information}
SM-1: Polymer characterizations. SM-2:  Model and Data for van der Waals interactions across polymer films supported by silica or oxidized silicon wafers. SM-3:  Secondary films do not originate from capillary effects. SM-4 : Free energy of free polymers in thin films: Contributions of entropy and van der Waals interactions.



\bibliography{biblio_paper3}

\end{document}